\documentclass[journal=jacsat,manuscript=article]{achemso}

\usepackage{hyperref}
\usepackage{hypcap}
\usepackage{textgreek}
\usepackage{graphicx}
\usepackage{soul} 

\usepackage{sidecap}
\usepackage{afterpage}
\usepackage{lineno}

\hypersetup{ 
colorlinks=true,
linkcolor=blue,
citecolor=blue
} 
\newcommand{\comm}[1]{}

\title{Fundamental Crystal Field Excitations in Magnetic Semiconductor SnO$_2$:Mn,Fe,Co,Ni}

\author{B. Leedahl}
\affiliation{Department of Physics and Engineering Physics, University of Saskatchewan, 116 Science Place, Saskatoon, Saskatchewan S7N 5E2, Canada}
\email{brett.leedahl@usask.ca}

\author{D. J. McCloskey}
\affiliation{Department of Physics and Engineering Physics, University of Saskatchewan, 116 Science Place, Saskatoon, Saskatchewan S7N 5E2, Canada}

\author{D. W. Boukhvalov}
\affiliation{College of Science, Institute of Materials Physics and Chemistry, Nanjing Forestry University, Nanjing 210037, P. R. China}
\alsoaffiliation{Institute of Physics and Technology, Ural Federal University, Mira 9 str., 620002 Yekaterinburg, Russia}

\author{I. S. Zhidkov}
\affiliation{Institute of Physics and Technology, Ural Federal University, Mira 9 str., 620002 Yekaterinburg, Russia}

\author{A. I. Kukharenko}
\affiliation{Institute of Physics and Technology, Ural Federal University, Mira 9 str., 620002 Yekaterinburg, Russia}

\author{E. Z. Kurmaev}
\affiliation{Institute of Physics and Technology, Ural Federal University, Mira 9 str., 620002 Yekaterinburg, Russia}
\alsoaffiliation{M.N. Mikheev Institute of Metal Physics of Ural Branch of Russian Academy of Sciences,
S. Kovalevskoi 18 str., 620990 Yekaterinburg, Russia}

\author{S. O. Cholakh}
\affiliation{Institute of Physics and Technology, Ural Federal University, Mira 9 str., 620002 Yekaterinburg, Russia}

\author{N. V. Gavrilov}
\affiliation{Institute of Electrophysics, Russian Academy of Sciences, Ural Branch, 620990 Yekaterinburg, Russia}

\author{V. I. Brinzari}
\affiliation{Pokatilov Laboratory of Physics and Engineering of Nanomaterials, Department of Theoretical Physics, Moldova State University, Chisinau MD-2009, Republic of Moldova}

\author{A. Moewes}
\affiliation{Department of Physics and Engineering Physics, University of Saskatchewan, 116 Science Place, Saskatoon, Saskatchewan S7N 5E2, Canada}
 
 \date{today}
 
\begin{document}

\begin{abstract}
Directly measuring elementary electronic excitations in dopant $3d$ metals is essential to understanding how they function as part of their host material. Through calculated crystal field splittings of the $3d$ electron band it is shown how transition metals Mn, Fe, Co, and Ni are incorporated into SnO$_2$. The crystal field splittings are compared to resonant inelastic x-ray scattering (RIXS) experiments, which measure precisely these elementary $dd$ excitations. The origin of spectral features can be determined and identified via this comparison, leading to an increased understanding of how such dopant metals situate themselves in, and modify the host's electronic and magnetic properties; and also how each element differs when incorporated into other semiconducting materials. We found that oxygen vacancy formation must not occur at nearest neighbour sites to metal atoms, but instead must reside at least two coordination spheres beyond. The coordination of the dopants within the host can then be explicitly related to the $d$-electron configurations and energies. This approach facilitates an understanding of the essential link between local crystal coordination and electronic/magnetic properties. 

\end{abstract}

\maketitle 

\section{Introduction}
SnO$_2$ is an important wide band gap ($\approx$\,3.6\,eV) oxide semiconductor with excellent optical transparency, metal-like conductivity, and high chemical stability, making it attractive for a myriad of applications.\cite{Zhou2014,Kim2002,Comini2002,Snaith2010} The observation of high temperature ferromagnetism in Co-doped SnO$_2$ by Ogale et al. in 2003 \cite{Ogale2003} induced a large number of experimental investigations of transition metal doped SnO$_2$.\cite{Lussier2004,Hays2005,Fitzgerald2006,Liu2007,Zhang2007,Yu2011,Li2011,Lamrani2014,Inpasalini2016}

Incidentally, high temperature ferromagnetism was also found also in SnO$_2$ doped with V, Cr, Mn, Fe and Ni. These studies revealed unquestionably that the properties of transition metal doped SnO$_2$ are highly sensitive to the sample synthesis conditions.\cite{Punnoose2005} A number of these experiments have suggested that the observed room temperature ferromagnetism is connected with the formation of oxygen vacancies. In one such study it was found that in SnO$_2$:Co free of oxygen vacancies, the Co atoms reside in a low spin state independent of concentration, and are ferromagnetic only through the exchange mechanism between nearest neighbour Co atoms.

However, if oxygen vacancies are introduced, they were found to reside near Co atoms, and strongly increase cobalt's magnetic moment, which depends heavily on the concentration and distribution of the dopant Co atoms.\cite{Wang2009} These vacancies can induce long range ferromagnetic order through the spin-split $3d^\uparrow$ and $3d^\downarrow$ impurity band exchange mechanism.\cite{Coey2005} In fact, oxygen vacancies in undoped SnO$_2$ were also found to produce ferromagnetism.\cite{Chang2005} Therefore, substituting metal ions of a different valency than the 4+ of Sn atoms in SnO$_2$ can assist in oxygen vacancy formation through the necessary charge compensation that must occur. This, in turn, would facilitate the appearance of ferromagnetism. 

Understandably, it is therefore of the utmost importance to understand how dopant atoms interact with their host material in order to validate their potential utility in electronic materials design. Herein, we present a method for comparing and displaying the crystal field excitations measured experimentally with those calculated theoretically.

\section{Experimental and Calculation Details}
Crystal field calculations performed with Quanty use the full atomic and crystal field multiplet theory and include the effects of the solid state.\cite{Haverkort2016} It accounts for the intra-atomic $3d$-$3d$ and $2p$-$3d$ Coulomb and exchange interactions, the atomic $2p$ and $3d$ spin-orbit couplings, and the crystal field splitting of the $e_g$ and $t_{2g}$ orbitals.\cite{Green2012,Haverkort2014} Quanty goes beyond density functional theory (DFT) based approaches that most often use a one electron basis.  Within the DFT framework, core-valence interactions can be modeled as an additional potential, but roughly speaking, the x-ray absorption spectrum is identical to the unoccupied density of states On the other hand, Quanty uses a local picture of the absorbing atom, and the solid is approximated by a small cluster in which all local Coulomb interactions and multiplets can be accounted for. This framework is necessary for many transition metal $L_{2,3}$ and rare earth $M_{4,5}$ edges, because the density of states is not what is observed in experiment, as the spectra are strongly influenced by multiplet effects.\cite{Leedahl2017_2} The resulting spectra are intrinsically many-body and should not be confused with the unoccupied density of states.

SnO$_2$ films were deposited on alumina substrate from 0.2 M alcohol solutions of SnCl$_4$ $\cdot$ 5H$_2$O via spray pyrolysis at temperatures in the interval 400-500$^\circ$C. The resulting thin film thicknesses were $\sim$\,100\,nm and were implanted with Mn, Fe, Co and Ni ions using a metal-vapor vacuum arc ion source. The operating pressure in the implantation chamber was kept below $2.0\times 10^{-2}$ Pa. The ion energy was set to 30 keV, and the pulse duration was 0.4 ms with a current density of 0.6 mA/cm$^2$, and an ion fluence of $2\times 10^{17}$ cm$^{-2}$. In order to ensure the lateral elemental homogeneity of the sample, a defocused ion beam with a diameter much larger than the sample size ($\approx$\,90\,mm versus 10\,mm) was used to dope the sample.  That is, sufficient measures we taken to eliminate the possibility of local elemental inhomogeneities. Furthermore, XAS and RIXS performed at two different beamlines, with different probing regions and areas produced consistent results; and XPS survey measurements were performed on three spatially different locations for each sample with similarly consistent results.

\begin{figure}
\begin{center}
\includegraphics[width=3.375in]{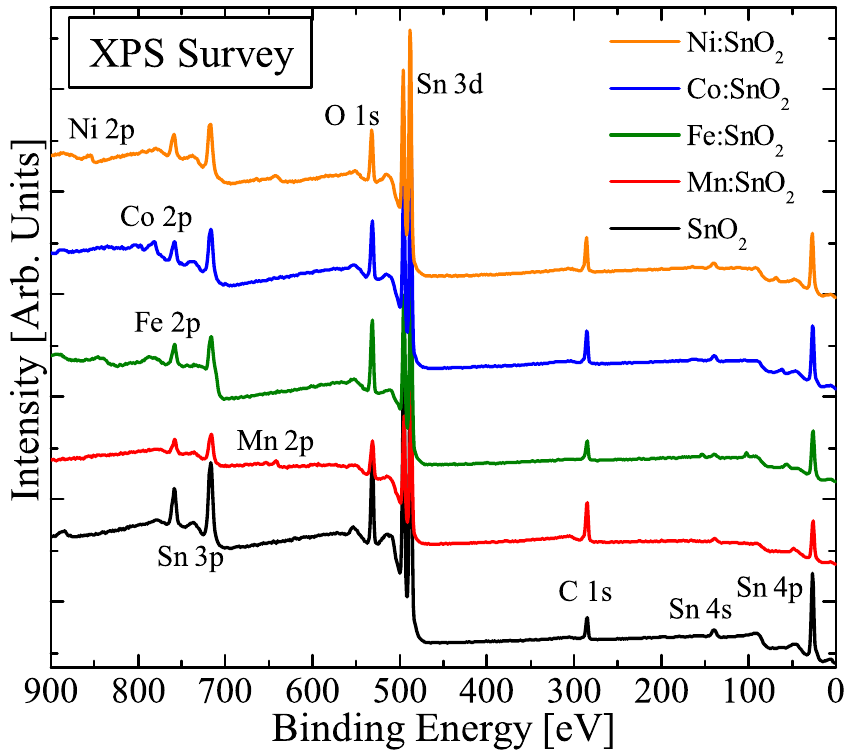}
\caption{Surface sensitive XPS survey spectra of $3d$-metal doped SnO$_2$ show that the samples are free of any major impurities, while showing some expected carbon contamination on the surface due to exposure to atmosphere.}
\label{fig:survey}
\end{center}
\end{figure}

X-ray photoelectron spectroscopy (XPS) measurements were made with an energy resolution of $\Delta$E $<$ 0.5 eV for Al K$\alpha$ radiation (1486.6 eV) and a beam spot size of 200 $\mu$m. The samples were held in vacuum (10$^{-7}$ Pa) for 24 hours prior to measurement, and only samples whose surfaces were free from micro impurities (as determined by means of surface chemical state mapping) were measured and reported herein.  The x-ray power load delivered to the sample was not more than 50 W in order to avoid x-ray stimulated decomposition of the sample. Under these conditions signal-to-noise ratios of at least 10000:3 were achieved in all cases. The XPS spectra were calibrated using the reference energy of 285.0 eV for the carbon $1s$ core level. The XPS survey spectra were measured for the binding energy range of 0-900 eV and are presented in Figure \ref{fig:survey}. No uncontrollable impurities were detected, which indicates the high quality of samples under investigation. The intensity of the dopant peaks relative to the Sn and O peaks of the XPS survey spectra were used to get an estimation of the dopant concentration within the probing depth of XPS,\cite{Biesinger2009} which is $\approx$\,5\,nm. Our analysis concluded that the average dopant concentration for all samples within this depth is about 4\%. This information, along with SRIM software calculations allowed us to compute an estimated dopant depth and concentration profile, which is plotted in the \emph{Supplemental Material}.\cite{Ziegler2010}

\begin{figure}
\begin{center}
\includegraphics[width=3.375in]{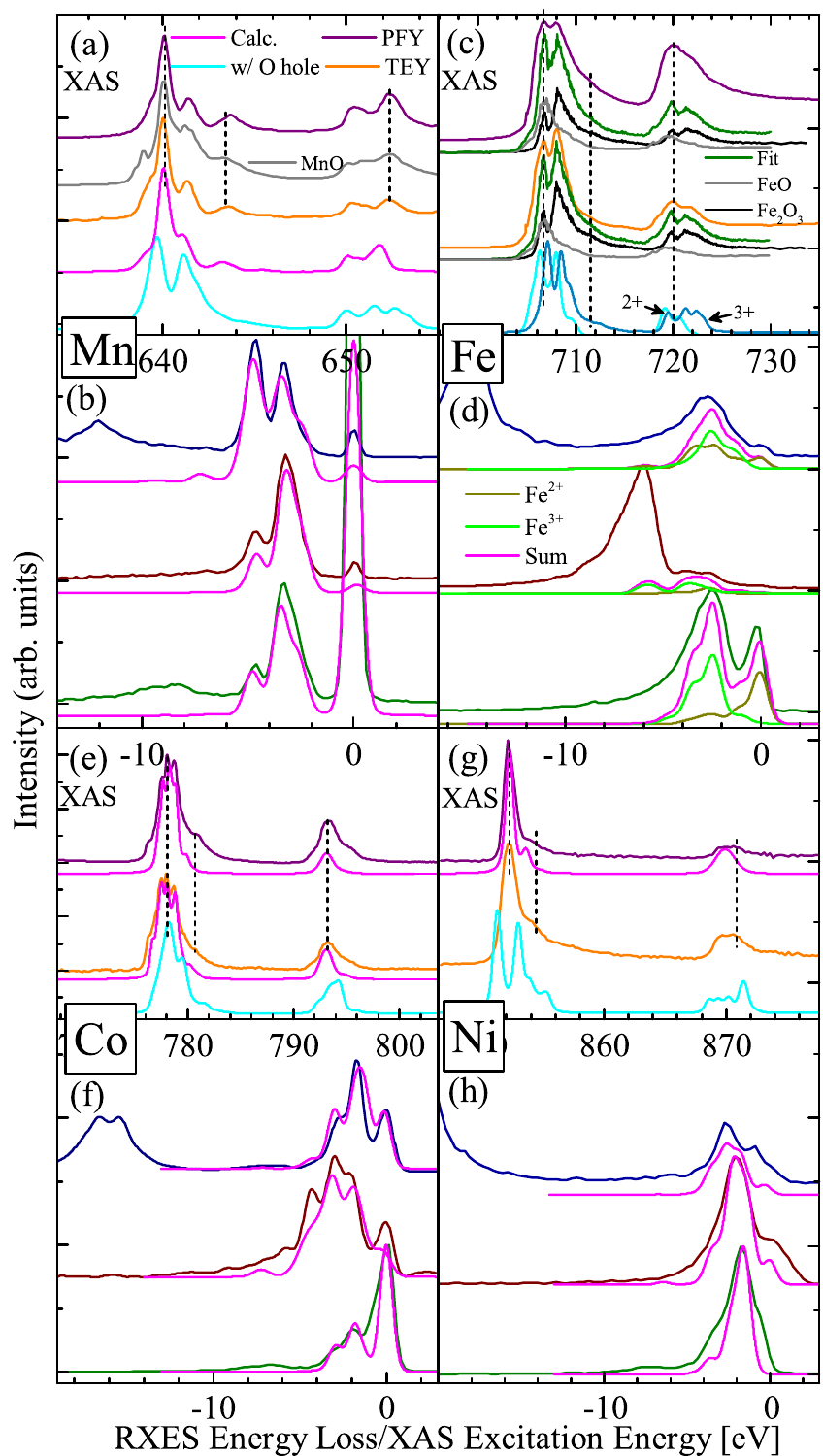}
\caption{The four quadrants of the figure show RIXS and XAS spectra for each Mn, Fe, Co, and Ni; the top panel in each quadrant displays the XAS spectra, while the bottom contains the RIXS. Crystal field calculations agreeing with the experimental spectra are shown in pink. The blue crystal field XAS calculations in panels (a), (c), (e), and (g) are of central importance. These calculations show what we would expect the experiment to resemble in the presence of a neighboring oxygen vacancy, which we do not see.}
\label{fig:rxesxas}
\end{center}
\end{figure}

X-ray absorption spectroscopy (XAS)\cite{Viswanatha2004} and resonant inelastic x-ray scattering (RIXS) experiments were performed at the REIXS beamline at the Canadian Light Source, and Beamline 8.0.1 at the Advanced Light Source, respectively. For these measurements the samples were oriented so the incident radiation was 45$^{\circ}$ to the surface normal, and the fluorescence CCD detector was fixed at 90$^{\circ}$ to the incident radiation. The incident photon beam was linearly polarized in the horizontal scattering plane, and count times for RIXS spectra ranged between 10 and 20 minutes. Given our experimental configuration, the attenuation length of x-rays in our energies of interest (640 to 870\,eV), relevant for photon-in photon-out techniques (RIXS and PFY), is between 60 and 100\,nm, which is on the scale of both the sample thickness ($\approx$\,100\,nm) and maximum dopant depth ($\approx$\,50\,nm), implying that the techniques herein are ideally suited to studies of such samples. XAS and RIXS spectra for each dopant are displayed in Figure \ref{fig:rxesxas}, where the panels (a), (c), (g), and (e) correspond to the XAS for a given dopant metal, while the lower panel in each quadrant shows the RIXS. Calculated spectra in this manuscript are always shown in pink.

\includegraphics[width=3.375in]{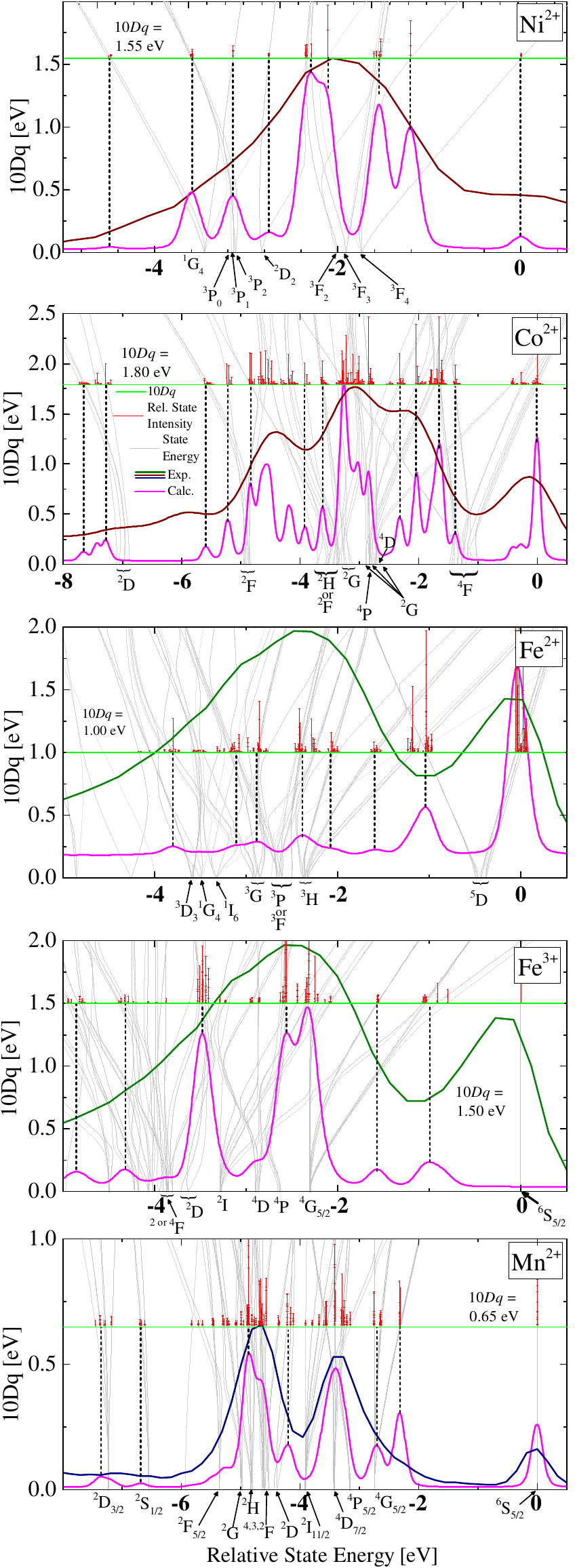}
\begin{figure}
\caption{Plotted in grey are Tanabe-Sugano diagrams (rotated 90$^\circ$) that show the individual state energies as grey lines as a function of $10Dq$ on the y-axis. Identical calculations from Figure \ref{fig:rxesxas}, but with significantly reduced broadening for clarity (pink) are also shown, along with the corresponding RIXS spectrum (the same spectra as in Figure \ref{fig:rxesxas}). As the individual energy states branch off and diverge as $10Dq$ increases, there comes a point in ``$10Dq$-space'' where the energy states agree with the experimental spectrum (this value of $10Dq$ is shown in green). Depending on the incident RIXS energy, these states will have various intensities. The calculated weighted intensities are shown in red, which overlap with the grey state energies at the correct $10Dq$ value. Upon broadening the red states the pink calculated spectra are produced, which in turn can be further broadened to match experiment. Term symbols in the form $^{2S+1}L_J$ are shown to illuminate the $d$-electron configurations from which each state originates.}%
\label{fig:TabSugAll}
\end{figure}%

\section{Results and Discussion}
\subsection{Crystal Field Excitations}
We have carried out a detailed analysis of the fundamental crystal field excitations present in $3d$ metals Mn, Fe, Co, and Ni that occur upon their implantation into the host SnO$_2$ semiconductor. The presence of a surrounding potential results in a crystal field splitting, which is observed as the breaking of degeneracy of $3d$ electron's energy levels.\cite{Ghiringhelli2005} By performing RIXS measurements we can directly access these energy levels. A RIXS spectrum---originating from crystal field effects---is determined by three things: the number of available electron configurations, their energy separations, and their weighted intensities as determined by transition probabilities. 

Shown in Figure \ref{fig:TabSugAll} are experimental spectra along with minimally broadened calculations beneath them (pink) to elucidate their underlying structure (note that the experimental spectra are the same as in Figure  \ref{fig:rxesxas}, as are the calculations, but with reduced broadening). That is, we have matched the experimental spectra to the broadened calculations to ensure that we can extract meaningful parameters; then in Figure \ref{fig:TabSugAll} we show the same calculations with reduced broadening to elucidate the underlying crystal field excitations. The calculations account for spectral intensity by computing the transition probabilities for each configuration. For the optimal value of the crystal field splitting parameter $10Dq$, where the calculation best matches experiment, a horizontal green line has been drawn. The background grey lines show the relative state energies that arise from, at a fundamental level, various $d$-electron configuration (i.e. $dd$ excitations). These state energies can be identified through term symbols in the form $^{2S+1}L_J$, which state the spin angular momentum ($S$), orbital angular momentum ($L$), and total angular momentum ($J$), of a given electron configuration. Labels for a chosen amount of these configurations are shown beneath the x-axis in each panel of Figure \ref{fig:TabSugAll}. It can be seen that as the crystal field (10Dq) varies, so do the relative energies of the various d-electron configurations. The crystal field parameters that result in calculated spectra most closely matching experimental RIXS data are marked with horizontal green lines. At this value of $10Dq$ the grey lines intersect with the the green line at energies that peaks appear in the experimental RIXS spectra. This allows us to visualize the nature of the crystal field that led to the experimental spectrum.

This is most easily observed for nickel in the top panel of Figure \ref{fig:TabSugAll}: at $10Dq = 1.55$ eV the state energies (grey) are such that upon calculation of their relative intensities (red), and then broadened to match experimental conditions, very good agreement is found with the experimental spectrum shown. This single value of $10Dq$ can then be used to describe \emph{all} XAS and RIXS spectra of Figure \ref{fig:rxesxas} (only one spectrum for each dopant is shown in Figure \ref{fig:TabSugAll} for clarity). Therefore,  precise analysis of the fundamental crystal field excitations that compose the experimental spectra can be performed. Essentially each peak or feature in a RIXS spectrum has its source in an elementary excitation. Through this methodology we can determine the basic electron configurations as identified through term symbols, as well as their relative intensities for any excitation energy. Meaning can then be assigned to each peak in an experimental RIXS spectrum. And because these energies are highly sensitive to the surrounding crystal lattice, fundamental information can be extracted regarding the nature of dopant metals in a host semiconductor. For example, how the local coordination of the dopant atoms relates to the available electronic configurations of the $d$-electrons, and their energy levels, which are important for functional uses.

For the other dopants, a similar analysis can be seen in Figure \ref{fig:TabSugAll}, wherein the source of all crystal field excitations can be determined. If no $J$ value is shown in the term symbol, it indicates several are present in that region and could not be clearly displayed. For cobalt, other crystal field parameters $Ds = -0.03$ and $Dt = 0.03$ are present, hence the non-degeneracy of term symbols (and hence energy levels) at $10Dq = 0$. In the case of Co$^{2+}$, the nature of the splitting of the ground $^4$F state implies that the $10Dq$ value can roughly be extracted from the energy separation between the first $dd$ excitations and the elastically scattered photons of the ground state.\cite{Wang2017} We observe that these do indeed occur at $\approx$ 1.8 eV below the 0 eV elastic peak. 

Considering iron, both 2+ and 3+ oxidation states are present and have been separated into these components in two panels of Figure \ref{fig:TabSugAll}. This makes the dissociation of crystal field excitations in the real material absolutely apparent to the eye. Each of the 2+ and 3+ Fe species contribute to overall spectral intensity at distinct energies. That is, the electron configurations of each oxidation state and their relative energies can be disentangled through this method, giving a more thorough insight to the spectral shape, and how each oxidation state provides its own energy levels to the material.

The well-known splitting of the two peaks in an Mn$^{2+}$ $L_2$ RIXS spectrum, which is a good measure of experimental resolution, can be identified as having originated from the splitting between the overall spin angular momentum configurations: the high energy peak is comprised mainly of $S = 3/2$ states and the low energy peak is entirely made up of $S=1/2$ configurations. 

This methodology can also be seen as a direct observation of Hund's rules. In all cases the ground state (at 0 eV) is the result of the highest spin state. And while the rules do not strictly apply beyond the ground state, they remain approximately true. As can be seen by the ordering of term symbols in each panel of Figure \ref{fig:TabSugAll}. As we move to higher energy states (right to left), the total spin angular momentum always decreases. Hund's second rule can similarly be seen (easiest in the Ni panel) in that the lower energy states have higher total orbital angular momentum where $L=0,1,2,3,4$ et cetera correspond to $L = S, P, D, F, G$ etcetera. And lastly, since the $3d$ shell is more than half full for Ni, Hund's third rule that the largest $J$ total angular momentum has the lowest energy can also be directly observed. In fact, within this picture, at non-zero values of the crystal field, we can see how Hund's rules become less applicable under the influence of a crystal field, by identifying the source of each individual RIXS peak in our spectra.

\subsection{Manganese}
Beginning with Mn dopants, the XAS spectra in Figure \ref{fig:rxesxas} (a) show good agreement between the bulk sensitive PFY and surface sensitive TEY, indicating identical incorporation of the dopant atoms in both regimes. Through crystal field calculations and the obvious similarity to MnO, the Mn ions in SnO$_2$ are entirely in a 2+ oxidation state. This is also evident from the $2p$ XPS spectra in Figure \ref{fig:2pxps} (a) in which the satellite peak at 646 eV in MnO---which must be used for comparison since the MnO and Mn$_2$O$_3$ $2p_{1/2}$ and $2p_{3/2}$ binding energies are too similar---is also very distinguishable in SnO$_2$:Mn. 

Given the quality of agreement between the crystal field calculations and the RIXS spectra in Figure \ref{fig:rxesxas} (b), we can be very confident in the nature of the local geometry of the Mn atoms as they reside in the host material. However, we can also be quite confident that the Mn atoms do not form bulk MnO by comparing their XAS spectra in Figure \ref{fig:rxesxas} (a) with each other. While they are similar, as one would expect from the similar octahedral environments in SnO$_2$ and MnO, they are clearly not identical. This is further supported by the crystal field parameters used to reproduce the experimental spectra. The crystal field splitting ($10Dq$) is 0.65 eV while the Slater integrals were reduced to 70\% of their Hartree-Fock values to account for the interacting of configurations like $d^5$ and $d^6\underline{L}$ not accounted for ($\underline{L}$ represents a hole in the ligand $2p$ band).\cite{Amorese2018} These configuration interaction processes reduce the intra-atomic electron-electron interactions as a result of the hybridization between metal $3d$ and oxygen $2p$ orbitals. 

This $10Dq$ value can be contrasted to the value usually calculated for MnO in the crystal field model, which is very close to 1.0 eV.\cite{Ghiringhelli2006} This difference is significant enough to conclude the Mn dopant atoms do not simply damage the host crystal upon implantation and form clusters of MnO, but rather do replace Sn atoms in their host environment. 

\subsubsection{Discussion of Oxygen Vacancies}
Furthermore---and this disussion applies to all dopants herein---charge neutrality must be maintained if 2+ ions are to replace Sn$^{4+}$ ions. However, we can confidently state that the local coordination of the dopant atoms (in all cases except cobalt) remains very nearly octahedral (as it is for Sn) because all crystal field parameters except $10Dq$ are equal to zero. Therefore, as is usually concluded with SnO$_2$, oxygen vacancies are likely present.\cite{Ahmed2017} 

This can also be understood from the perspective of the compatibility of ionic radii between the host crystal and the dopants. If we consider the ionic radii of the 6-coordinated elements, we find that the dopants in a 2+ oxidation state match well with that of Sn$^{4+}$, which is 0.83\,{\AA}. The most similar ionic radii for each of Mn, Co, and Ni is for the 2+ low-spin configuration (0.81\,{\AA}, 0.79\,{\AA}, and 0.83\,{\AA}, respectively). This is exactly the overwhelmingly observed oxidation state present in each of the doped samples according to our measurements. For iron, the situation is slightly different in that low-spin Fe$^{2+}$ has an ionic radius of 0.75\,{\AA}, while high-spin Fe$^{3+}$ is slightly closer to that of Sn$^{4+}$, with an atomic radius of 0.79\,{\AA}, and so there is a tendency towards Fe$^{3+}$ incorporation in SnO$_2$.\cite{Coey2004} Nonetheless, this is actually quite consistent with our XAS and XPS measurements, which indicate that both 2+ and 3+ oxidation states are present for Fe doped SnO$_2$. That is, from the simplified view only considering the compatibility of atomic radii, we obtain a good first order approximation as to \emph{why} oxygen vacancies are formed at all. Because of the mismatch in cation charge between the host and the dopant, the total charge must be compensated for by the formation of oxygen vacancies somewhere in the sample.

However, of central importance to this manuscript is to show that through our XAS spectra we can rule out the possibility that the oxygen vacancies exist as nearest neighbours to the metal atoms.\cite{Erdem2010} This is in contrast to the findings of previous theoretical approaches, in which they calculate that oxygen vacancies are heavily favored as nearest neighbours to the dopant metal ion.\cite{Wang2009_2} Using our techniques, the presence of nearest neighbour oxygen vacancies, if it were to occur throughout the crystal in any meaningful proportion, would heavily distort the measured spectra, and also the crystal field parameters used to model them.\cite{Green2015} The degree to which this distortion would occur can be calculated by determining how the crystal field parameters would change in the presence of a nearest neighbour oxygen vacancy. This is accomplished by calculating the Madelung potential on the central atom due to the surrounding charge distribution. As it turns out, the ratio of the crystal field parameters for the metal atom would change significantly if an oxygen vacancy were neighbouring it, to a ratio of $Dq:Ds:Dt = 1.00:0.44:0.03$. Such spectra, when calculated and compared to experiment (blue calculations in Figure \ref{fig:rxesxas}), show very poor agreement; and in the case of ion implanted SnO$_2$ it is apparent that such vacancies do not exist. That is, we are able to rule out the existence of nearest neighbour oxygen vacancies to the metal atoms by calculating what the XAS spectrum would look like if there were nearest neighbour oxygen vacancies. By showing the obvious disagreement, we can conclude that this situation does not manifest itself in our thin films.

\subsection{Iron}
The situation with Fe ions in SnO$_2$ is more complicated; there exists a combination of 2+ and 3+ Fe ions (see XPS in Figure \ref{fig:2pxps} (b)), but the ratio between the two varies significantly between the surface and bulk regimes, similar to what has been found in Bi$_2$Te$_3$.\cite{Leedahl2017} To determine this ratio, the standard oxides FeO and Fe$_2$O$_3$ were used as component spectra, and added in a linear combination to produce the best fit to the SnO$_2$:Fe TEY and PFY spectra. The component spectra and their sums can be seen in Figure \ref{fig:rxesxas} (c). The PFY (bulk) spectrum is fit with 60\% Fe$_2$O$_3$ and 40\% FeO, while the TEY (surface) spectrum is fit with 75\% Fe$_2$O$_3$ and 25\% FeO. This result is consistent with what is found using a linear sum of calculated 2+ and 3+ spectra for the RIXS as well. Figure \ref{fig:rxesxas} (d) displays the component calculations as well as the overall sum of the two. The agreement with the $dd$ excitations of the experiment is quite good, while the charge transfer excitations at higher energy losses are unaccounted for in crystal field theory.

The calculation parameters were the following: $10Dq$ = 1.0 eV with a 30\% reduction of the Slater integrals for Fe$^{2+}$; and $10Dq$ = 1.5 eV with a 50\% reduction of the Slater integrals for Fe$^{3+}$. All other crystal field parameters were set to 0 eV, indicating that both oxidation states of iron are incorporated into the lattice in an octahedral symmetry, not noticeably distorting the host structure. And similar to Mn, Fe was not found to exist in nearest-neighbor proximity to oxygen vacancies.

\subsection{Cobalt}
The XPS measurements in Figure \ref{fig:2pxps} (c) confirms our XAS results that Co is completely in a 2+ state. The non-zero crystal field $Ds$ and $Dt$ symmetry parameters of Co in SnO$_2$ that led to our best fit suggest that a significant distortion is present local to these dopant atoms. For the RIXS spectra it was found that $10Dq = 1.8$ eV, $Ds = -0.03$ eV, $Dt = 0.03$ eV, and the Slater integrals scaled to 67\% of their nominal values produced the best fit. This distortion is actually larger than the difference between perfect octahedral symmetry and the near octahedral symmetry found in SnO$_2$. In nominal rutile SnO$_2$ the six O ligands surrounding the Sn atoms are not all at 90$^\circ$ as would be the case in perfect octahedral symmetry; but rather the six O-Sn-O angles are 78$^\circ$,78$^\circ$, 102$^\circ$, 102$^\circ$, 90$^\circ$, and 90$^\circ$. 

As was mentioned above, calculating the Madelung potential on the central Sn atom yields a good estimate of how the crystal field parameters vary in response to slight distortions in perfect octehedral symmetry.. In this case the rutile structure would vary the ratios of the crystal field parameters $10Dq:Ds:Dt:Du:Dv$ from 1.0 : 0 : 0 : 0 : 0 in perfect octahedral symmetry to 1.0 : -0.005 : -0.005 : -0.011 : -0.040. As it turns out, the difference between $O_h$ and rutile calculation is too small to be experimentally detected within our resolution. This is the reason the other dopants can be sufficiently described purely by a $10Dq$ value alone, without having to resort to non-zero lower symmetry parameters. However, the non-zero -0.03 parameters of $Ds$ and $Dt$ required to obtain adequate agreement with experiment here is significant enough to be detected. Therefore, the crystal geometry in the case of SnO$_2$:Co is measurably more warped than for the other dopants studied herein. This suggests that oxygen vacancies tend to reside nearer to the Co atoms as compared to the other dopants studied herein; and it is likely that in some significant fraction the oxygen vacancies are nearest neighbours to the Co atoms.

\subsection{Nickel}
\begin{figure}
\begin{center}
\includegraphics[width=3.375in]{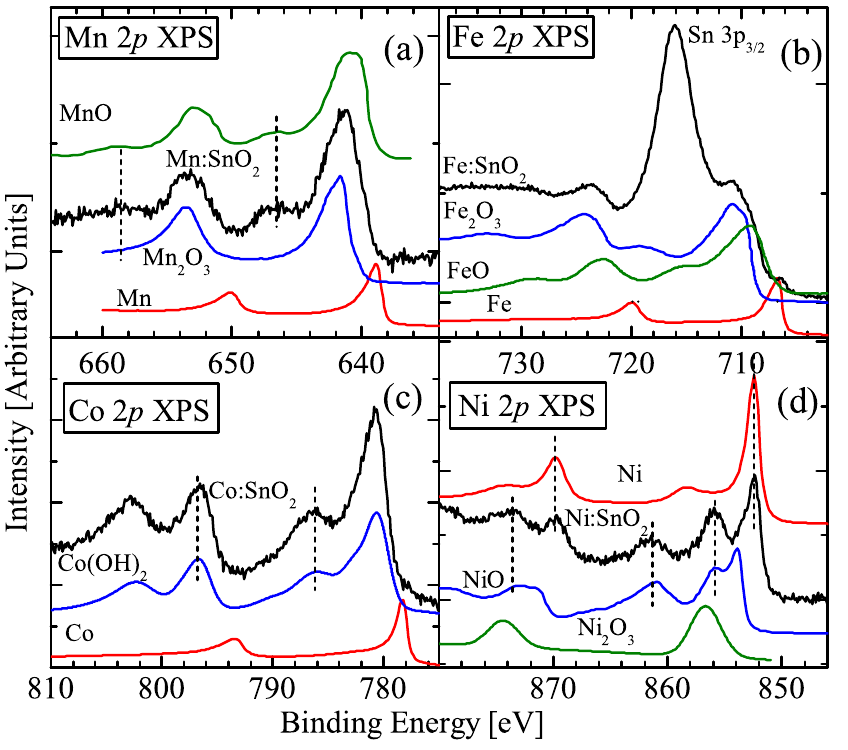}
\caption{The binding energies of metal $2p$ electrons measured with XPS can be used to determine the oxidation states present near the surfaces, and well as their approximate proportions. In all cases except iron (3+), 2+ metal ions dominate the surface states.}
\label{fig:2pxps}
\end{center}
\end{figure}

As is usually the case for Ni, it is primarily found in a 2+ oxidation state, but the XPS spectra of Figure \ref{fig:2pxps} (d) and the broadened nature of the TEY in Figure \ref{fig:rxesxas} (g) suggest there is some metallic clustering, but only on the surface layers.

In the bulk, the situation with nickel is somewhat unique in that the crystal field splitting is significantly larger as compared to its role in its most common oxide, NiO, which is generally calculated to be $10Dq=$1.05 eV.\cite{Huotari2008} To reproduce the spectra of SnO$_2$:Ni, a $10Dq$ value of 1.55 eV was used, while reducing the Slater integrals to 50\% of their nominal values. These values suggest that there is a rather large degree of covalent bonding between the octahedrally coordinated Ni atoms and surrounding O atoms. 

The shorter bond lengths in SnO$_2$ as compared to NiO is consistent with the higher crystal field splitting we would expect if Ni atoms are substituted into Sn sites with $O_h$ coordination. In fact, it has previously been shown that the crystal field value in NiO depends quite strongly on the local bond length.\cite{Drickamer1973, Barreda2017} This reduction in space for the Ni ions would result in a stronger crystal field splitting, as well as an increased orbital overlap with the ligand atoms, which we indeed see in our model as the largely reduced Slater integrals.

\section{Conclusions}
Understanding RIXS lineshapes is a powerful tool for understanding a material's basic properties. Given that there is no value to a spectrum unless some insight can be extracted from it, we show that by precisely comparing the contributions of individual crystal field excitations to experiment, one can ascertain more knowledge that was previously unavailable. In essence, we can answer the question of why spectra have the shapes we observe. The individual excitations have various intensities, which are a function of input experimental photon excitation energy, and can be calculated within a crystal field model. Once the weighted state intensities are broadened to match experiment, a direct comparison between theory and experiment can be made to gather insight on the fundamental electron excitations that are crucial for many of the properties of condensed matter systems. For transition metal doped SnO$_2$ this methodology provides a means of observation that leads us to conclude that via the synthesis method used herein, oxygen vacancy formation cannot reside next to the dopant atoms.

Furthermore, our findings suggest that there is substantial potential for future studies, utilizing a wide variety of experimental and theoretical techniques to further understand the link between the structure, electronic, and magnetic properties. For example, a comprehensive density functional theory study could be useful to explore the magnetic effects of vacancy pair distance, vacancy concentration, and vacancy coordination, as well as the relevant exchange interactions that give rise to the magnetic moment.\cite{Aravindh2015,Espinosa2011}

\section{Acknowledgments}
The XAS measurements were performed at the Canadian Light Source, and XES measurements were performed at the Advanced Light Source, both were supported by the Natural Sciences and Engineering Research Council of Canada (NSERC) and the Canada Research Chair program. The ion implantation of SnO2 thin films is performed under support of the Republic of Moldova (project no. 15.817.02.29F). The XPS measurements are supported by the Russian Science Foundation for Basic Research (Project 17-08-00395), the Ministry of Education and Science of Russia (State task of 3.7270.2017/8.9) and by the Ural Branch, Russian Academy of Sciences (project no. 18-10-2-6, multidisciplinary program).

\providecommand{\latin}[1]{#1}
\makeatletter
\providecommand{\doi}
  {\begingroup\let\do\@makeother\dospecials
  \catcode`\{=1 \catcode`\}=2 \doi@aux}
\providecommand{\doi@aux}[1]{\endgroup\texttt{#1}}
\makeatother
\providecommand*\mcitethebibliography{\thebibliography}
\csname @ifundefined\endcsname{endmcitethebibliography}
  {\let\endmcitethebibliography\endthebibliography}{}

\begin{figure}
\begin{center}
\includegraphics[width=3.25in]{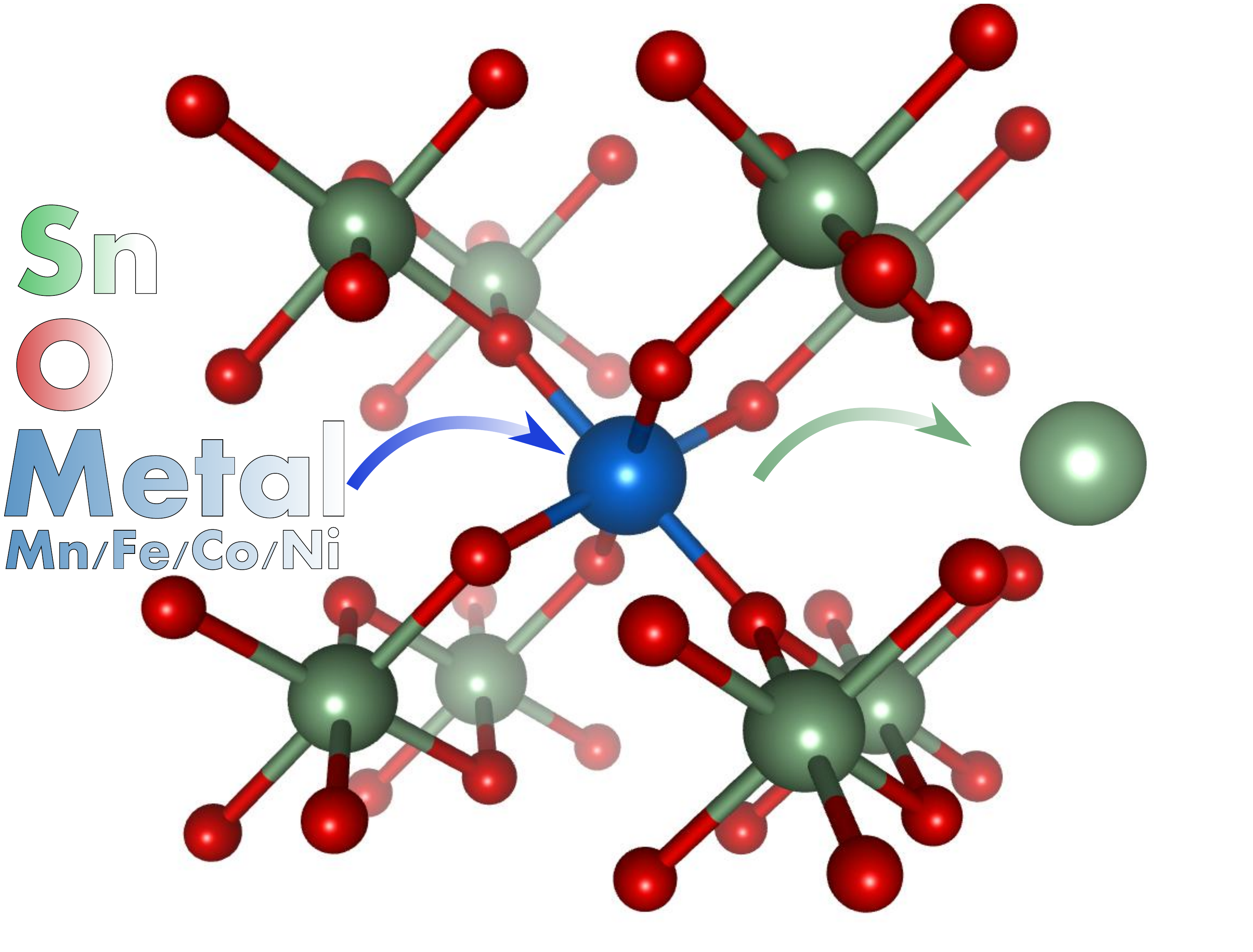}
\caption{Table of Contents graphic}
\label{fig:TOC}
\end{center}
\end{figure}

\end{document}